\documentclass[prb,twocolumn,superscriptaddress,showpacs]{revtex4}
\usepackage{bm}
\usepackage{amsmath}
\usepackage{amssymb}
\usepackage[dvips]{graphicx}
\begin{document}

\title{Resonant polaron-assisted tunneling of strongly interacting electrons through single-level vibrating quantum dot}

\author{Gleb A. Skorobagatko }

\email{gleb_skor@mail.ru}
\affiliation{ B.Verkin Institute for Low Temperature Physics \&
Engineering,  National Academy of Sciences of Ukraine, 47 Lenin
Ave, Kharkov, 61103, Ukraine}

\pacs{73.63.Kv, 72.10.Pm, 73.23.-b}

\begin{abstract}
The problem of resonant transport of strongly interacting electrons through a one-dimensional single-level vibrating quantum dot is being considered. In this paper, we generalize the Komnik and Gogolin model \cite{KG1} for the single-electron transistor with $g=1/2$- Luttinger liquid leads to the case of strong electron-vibron interaction in a quantum dot. The effective transmission coefficient and differential conductance of the system has been derived for the general case of asymmetric tunnel barriers. The main result obtained is that, in the zero-temperature limit, the resonant polaron-assisted tunneling with perfect transmission is possible. This resonant tunneling is of the novel (Andreev-like) type due to a special electron-electron interaction in the leads. As a result, a strong domination of resonant polaron-assisted electron transport at low temperatures has been found. Additional narrowing due to electron-electron interaction in the leads, is roughly the same for all polaron-assisted resonances.
\end{abstract}

\maketitle

\section{Introduction}
The problem of electron transport through various molecular structures and, particularly, through molecular single-electron transistors (SETs)has become a hot topic in the modern mesoscopic physics \cite{PP,PT,NR,GR}. A single-electron transistor in question is modelled as a single-level quantum dot (QD), weakly coupled to two one-dimensional leads (quantum wires or carbon nanotubes) \cite{FUR,FL,MCK,KG1,KG2,OUR}. One can control the tunnel current through a described system by means of two independent parameters: (i) "driving" voltage (bias) $V$ between the leads and, (ii) "gate" voltage $V_{g}$, which is able to move the fermionic level of the QD. Besides that, the average current depends on the properties of tunnel barriers, on the electron-vibron coupling in the QD, and on the strength of electron-electron interaction in the system. For the most general case of arbitrary electron-electron and electron-vibron interactions, it is impossible to solve the transport problem exactly. Nevertheless, some limiting cases are solvable. 

It is widely known \cite{MCK,FL,GS,KO,OUR,NEW}, that the vibrations of QD qualitatively change the character of low-temperature electron transport through a SET even with Fermi-leads. Two main effects characterize the low-temperature differential conductance of the system with vibrating QD in the case of Fermi leads: (i) vibron-assisted tunneling, where the additional (satellite) resonant peaks of differential conductance emerge \cite{MCK,FL,GS}, and (ii) the polaronic narrowing (Frank-Condon blockade) of the widths of all resonances \cite{MCK,KO}(for strong enough electron-vibron coupling). 

One-dimensional leads imply electron-electron interaction in the system, which is described by Luttinger liquid model with the dimensionless correlation parameter $g$ ($0<g<1$). In the case of strong electron-electron interaction ($g<1/2$), perturbation theory calculations in "bare" level width of the fermionic level of QD are valid even at low temperatures (i.e. the sequential tunneling limit is maintained at all temperatures). This results in the strong suppression of tunneling probabilities, due to the well-known Kane-Fisher effect \cite{KF,FUR} and in the vanishing of the conductance at $T=0$. This is not the case for weak electron-electron interaction ($1\geq g\geq 1/2$), where one could observe a resonant regime of electron transport at low temperatures in the case of symmetric tunnel barriers \cite{KF,NG}. The consequence of this fact is the perfect ($G_{0}=e^{2}/h$) zero-temperature conductance at $V\rightarrow 0$\cite{KF,NG}. 

Thus, it is interesting to understand, how the interplay of the effects of strong electron-electron and electron-vibron interactions influences electron transport through a single-level QD. Although some results in this direction have been already obtained for the sequential tunneling limit \cite{OUR,CHINA}, these considerations are unable to give us correct predictions for the special $g=1/2$-case. It was shown \cite{KF,KG1} that in the absence of electron-vibron interaction, a possibility of resonant transport of strongly interacting electrons (with conductance quantum at $T\rightarrow 0$ for the symmetric case) still exists for $g\geq 1/2$. At the special value $g=1/2$ of LL correlation parameter the problem of electron transport in such a strongly interacting system is reduced to the scattering problem for the noninteracting fermions, which is exactly solvable \cite{KG1,KG2}. Furthermore, it is impossible to obtain this solution by any perturbative method. The origin of this fact lies in the special type of symmetry in the correlations between the electrons from different leads, which is specific only for $g=1/2$-system. The generalization of the $g=1/2$- model \cite{KG1,KG2} to the case of weak electron-vibron coupling is proposed in Ref.[\onlinecite{K}]. The solution of Ref.[\onlinecite{K}] treats weak electron-vibron coupling perturbatively. Thus, it does not describe the effects of resonant polaron-assisted tunneling, which take place at sufficiently strong electron-vibron coupling. Therefore, a problem of resonant tunneling for $g=1/2$-model with strong electron-vibron interaction still needs further considerations. 

Below, we consider the influence of quantum vibrations on resonant electron transport through a single-level QD, weakly coupled to the $g=1/2$- Luttinger liquid leads. In the polaronic approximation analytical formulae are derived  for the
effective transmission coefficient and differential conductance. The solution describes both the resonant tunneling regime for strong electron-vibron interaction and the sequential tunneling. In particular, we have reproduced a known nonperturbative result for elastic tunneling $G(V\rightarrow 0, T\rightarrow 0)=G_{0}=e^{2}/h$ for symmetric junction. Interestingly, this is not the case for general (asymmetric) QD, where a Kane-Fisher suppression of conductance peak at zero bias still takes place. In the strong coupling regime, a correct description of all "polaron-assisted" resonances, appearing at energies $\varepsilon_{l}=\hbar\omega_{0}l$, ($l=\pm1,2,..$) is obtained. At the zero-temperature limit, the amplitude of each polron-assisted resonant peak in symmetric junction reaches conductance quantum, signalling a perfect transmission of polaron at resonance energies. We show that the widths of resonances behave non-monotonically, as the functions of electron-vibron interaction constant. Moreover, the details of such behaviour in our model differ from the predictions of a similar model with noninteracting leads \cite{NEW} ($g=1$ case). Particularly, an additional narrowing of widths of all polaron-assisted resonances induced by electron-electron interaction is found. A physical interpretation of such narrowing consists in the special type of strong correlations between electrons from different leads, which affect the tunneling by the additional (polaron-assisted) channels. Thus, one can distinguish the resonant polaron-assisted tunneling  in the interacting ($g=1/2$) and noninteracting ($g=1$) systems.

\section{The model}
The Hamiltonian of our model reads
\begin{equation} \label{1}
\hat{H}=\sum_{j=L,R}\hat{H}_{l}^{(j)}+\hat{H}_{d}+\sum_{j=L,R}
\hat{H}_{t}^{(j)}.
\end{equation}
In Eq.(1) the first term describes the Hamiltonian of one-dimensional infinite Luttinger liquid leads, which is quadratic in the bosonized form (we consider the spinless electrons) $\hat{H}_{l}^{(j)}= 1/4 \pi \int dx(\partial_{x}\Phi_{j}(x))^{2}$. (Here and below we put $\hbar v_{g}=\hbar v_{F}/g=1$ with "bare" Fermi velocity $v_{F}$.) The bosonic phase fields 
$\Phi_{j}(x)$ of the $j$-th lead ($j=L,R$) are connected with corresponding field operators $\hat{\Psi}_{j}(x)$ of chiral fermions on the $j$-th lead in strongly nonlinear way by the standard bosonization formula 
$\hat{\Psi}_{j}(x)=\exp(i\Phi_{j}(x)/\sqrt{g})/\sqrt{2\pi a_{0}}$ (where $a_{0}$ is the corresponding lattice constant) \cite{KG1}. The chiral charge density operator $\hat{\rho_{j}}(x)$ at the point $x$ of each $j$-th lead can be performed in both fermionic and bosonic representations as follows 
$\hat{\rho_{j}}(x)=\hat{\Psi}_{j}^{+}(x)\hat{\Psi}_{j}(x)=\partial_{x}\Phi_{j}(x)/2\pi\sqrt{g}$. Here $g$ - is the dimensionless Luttinger liquid correlation parameter. Fermionic field operator $\hat{\Psi}_{j}(x)$ describes the annihilation of chiral fermion, living in one-dimensional infinite $j$-th channel ($j=L,R$), corresponding to the $j$-th physical lead of the system. However, a negative half-axis $x<0$ of each $j$-th channel corresponds to the incoming particles moving from the infinity towards the boundary (i.e. to the point $x=0^{-}$), while the positive one stands for  the particles, which move back from the boundary (i.e. from the point $x=0^{+}$) to the infinity in the $j$-th channel. The following relation is defined at the boundary (in the vicinity of the point $x=0$ of each $j$ channel) \cite{KG1} $\hat{\Psi}_{j}(0)=(\hat{\Psi}_{j}(0^{-})+\hat{\Psi}_{j}(0^{+}))/2$. It is evident from the above and due to the continuity of chiral electric charge distribution along the system, that the average current through the system (i.e. from the left electrode to the right one or vice versa) can be performed in a very transparent form in terms of chiral density operators $\hat{\rho_{j}}(x)$ in the arbitrary (left or right) lead \cite{KG2}
\begin{equation} \label{2}
\bar{I}=\langle \hat{I}_{L(R)} \rangle=(e/\hbar)\{ \langle \hat{\rho}_{L(R)}(0^{-}) \rangle - \langle \hat{\rho}_{L(R)}(0^{+}) \rangle \}.
\end{equation}
The second term of Eq.(1) describes the Hamiltonian of single-level quantum dot (QD) coupled to a quantum harmonic oscillator which models the vibrations of QD. Following Refs.[\onlinecite{KG1,KG2}] we also take into account Coulomb interaction between the leads and QD.
 \begin{eqnarray}\label{3}
\nonumber
\hat{H}_{d}=\{\Delta \hat{d}^{+}\hat{d}+\lambda_{C}\hat{d}^{+}\hat{d}\sum_{j=L,R}\hat{\Psi}_{j}^{+}(0)\hat{\Psi}_{j}(0)\}+\nonumber\\
+\hbar\omega_{0}\{\frac{1}{2}(\hat{p}^{2}+\hat{x}^{2})+\lambda \hat{x}\hat{d}^{+}\hat{d}\}.
\end{eqnarray}
In Eq.(3) $\hat{d}^{+}(\hat{d})$ - are the standard fermionic creation (annihilation) operators 
($\lbrace\hat{d},\hat{d}^{+}\rbrace=1$) for the electron on QD, $\Delta$ - is the level energy, $\hbar\omega_{0}$ is the
energy of vibrational quantum (here $\omega_{0}$ is the self frequency of quantum oscillator) and $\hat{p}$,$\hat{x}$ are the dimensionless bosonic operators of the momentum and center-of-mass coordinate of QD. The interaction in the QD is controlled by two independent parameters: (i) $\lambda$ is the dimensionless electron-vibron coupling constant, and (ii) $\lambda_{C}$ is the dimensionless constant of Coulomb interaction. In our model we have fixed $\lambda_{C}$ at so called Toulouse point ($\lambda_{C}=2\pi$), while $\lambda$ remains as a free parameter (below we will be interested mostly in the case of strong electron-vibron coupling $\lambda\geq1$). The last term in Eq.(1) represents the tunneling Hamiltonian
\begin{equation} \label{4}
\hat{H}_{t}^{(j)}=(\gamma_{j}\hat{d}^{+}\hat{\Psi}_{j}(0)+h.c.).
\end{equation}
Here $\gamma_{j}$ are the tunneling amplitudes. 

The Hamiltonian $\hat{H}$ of Eqs.(1-4) can be transformed into a more convenient form $\hat{\tilde{H}}$ by introducing new symmetric and antisymmetric phase fields $\Phi_{\pm}(x)=(\Phi_{L}(x)\pm\Phi_{R}(x))/2$ and by applying of two commuting unitary transformations $\hat{U}_{f}=\exp[-i(\hat{d}^{+}\hat{d}-1/2)\Phi_{+}(0)/\sqrt{2g}]$ and 
$\hat{U}_{b}=\exp(-i\lambda\hat{p}\hat{d}^{+}\hat{d})$, $\hat{\tilde{H}}=\hat{U}_{f}\hat{U}_{b}\hat{H}\hat{U}_{b}^{-1}\hat{U}_{f}^{-1}$.
Here the transformation $\hat{U}_{b}$ removes the electron-vibron interaction term from the Hamiltonian (3) of the QD. As it was shown \cite{KG1}, by applying unitary transformation $\hat{U}_{f}$ to the Hamiltonian (1) at $g=1/2$, one can rewrite it in terms of new fermions $\hat{\Psi}_{\pm}(x)=\exp(i\Phi_{\pm}(x))/\sqrt{2\pi a_{0}}$ and remove the $\Phi_{+}(x)$- phase field from the tunneling term \cite{KG1}. At the Toulouse point $\lambda_{C}=2\pi$ one can remove also the Coulomb interaction term from the transformed Hamiltonian of QD. Thus, if $g=1/2$ and $\lambda_{C}=2\pi$, one could rewrite the total transformed Hamiltonian in the form $\hat{\tilde{H}}=\hat{\tilde{H}}_{l}+\hat{\tilde{H}}_{d}+\hat{\tilde{H}}_{t}$, where $\hat{\tilde{H}}_{l}=\sum_{\pm}1/2\pi\int dx(\partial_{x}\Phi_{\pm}(x))^{2}$ and
\begin{equation} \label{5}
\hat{\tilde{H}}_{d}=\tilde{\Delta}\hat{d}^{+}\hat{d}+\frac{\hbar\omega_{0}}{2}(\hat{p}^{2}+\hat{x}^{2})
\end{equation}
are quadratic now (here $\tilde{\Delta}=\Delta-(\lambda^{2}/2) \hbar\omega_{0}$).
The transformed tunneling Hamiltonian takes the form
\begin{eqnarray}\label{6}
\nonumber
\hat{\tilde{H}}_{t}=\hat{d}^{+}\hat{X}^{+}[\gamma_{L} \hat{\Psi}_{-}(0)+\gamma_{R} \hat{\Psi}_{-}^{+}(0)]+\nonumber\\
+[\gamma_{L} \hat{\Psi}_{-}^{+}(0)+\gamma_{R} \hat{\Psi}_{-}(0)]\hat{X}\hat{d}.
\end{eqnarray}
Here the operator $\hat{X}=\exp(i\lambda\hat{p})$ describes the influence of electron-vibron interaction on tunneling, while the coupling terms $\hat{d}^{+}\hat{X}^{+}\gamma_{R}\hat{\Psi}_{-}^{+}(0)$ and $\gamma_{R}\hat{\Psi}_{-}(0)\hat{X}\hat{d}$ reveal the existence of additional Andreev-like tunneling of $\hat{\Psi}_{-}$-fermions. Operators $\hat{\Psi}_{\pm}(x)$ stand for new fermions and fulfill standard fermionic anticommutation relations in the Schroedinger representation $\{\hat{\Psi}_{\pm}(x),\hat{\Psi}_{\pm}^{+}(x^{'})\}=\delta(x-x^{'})$. It is evident from the above, that new fermionic field operators $\hat{\Psi}_{\pm}(x)$($\hat{\Psi}_{\pm}^{+}(x^{'})$) annihilate (create) a new nonlocal fermion, which exists simultaneously on both (left and right) physical leads of the system on the distance $\vert x \vert$ from the QD (i.e. from the point $x=0$) \cite{KG1,KG2}. Thus, special electron-electron interaction in the system entangles real electrons from different (left and right) physical leads in a sufficiently nonlinear way, making them strongly correlated. Density operators $\hat{\rho}_{+}(x)$ and $\hat{\rho}_{-}(x)$ for $\hat{\Psi}_{\pm}$ - fermions $\hat{\rho}_{\pm}(x)=\hat{\Psi}_{\pm}^{+}(x)\hat{\Psi}_{\pm}(x)=\partial_{x}\Phi_{\pm}(x)/\sqrt{2}\pi$ define chiral charge- ($\hat{\rho}_{+}(x)$) and current-($\hat{\rho}_{-}(x)$) densities measured in the symmetric points on the leads on the distance 
$\vert x \vert$ from the QD, $\hat{\rho}_{\pm}(x)=\hat{\rho}_{L}(x)\pm\hat{\rho}_{R}(x)$. Since $\Phi_{+}(x)$ - channel is decoupled now, one can represent the average current as follows \cite{KG1,KG2}
\begin{equation} \label{7}
\langle \hat{I} \rangle=(e/\hbar)\{\langle \hat{\Psi}_{-}^{+}\hat{\Psi}_{-}(0^{+}) \rangle - \langle \hat{\Psi}_{-}^{+}\hat{\Psi}_{-}(0^{-}) \rangle\}.
\end{equation}

\section{QEM-method and fermion-boson factorization}

To solve the described model we use a well-known quantum equation of motion (QEM) method. The Heisenberg equations for the fermionic operators take the form
\begin{equation} \label{8}
i\hbar\partial_{t}\hat{d}=\tilde{\Delta}\hat{d}+\hat{X}^{+}[\gamma_{L} \hat{\Psi}_{-}(0)+\gamma_{R} \hat{\Psi}_{-}^{+}(0)]
\end{equation}
where, following Refs.[\onlinecite{KG1,KG2}] we defined
$\hat{\Psi}_{-}(0)=(\hat{\Psi}_{-}(0^{-})+\hat{\Psi}_{-}(0^{+}))/2$, and
\begin{equation} \label{9}
i\hbar\partial_{t}\hat{\Psi}_{-}(x)=-i\partial_{x}\hat{\Psi}_{-}(x)+\delta(x)[\gamma_{L}\hat{X}\hat{d}-\gamma_{R}\hat{X}^{+}\hat{d}^{+}],
\end{equation}
here $\delta(x)$ is the delta function. Integrating Eq.(9) in the vicinity of point $x=0$, one obtains
\begin{equation} \label{10}
i[\hat{\Psi}_{-}(0^{+})-\hat{\Psi}_{-}(0^{-})]=\gamma_{L}\hat{X}\hat{d}-\gamma_{R}\hat{X}^{+}\hat{d}^{+}.
\end{equation}
In the absence of electron-vibron coupling ($\hat{X}^{+}=\hat{X}=1$), Eqs.(8-10) are reduced to Eq.(6) from Ref.[\onlinecite{KG1}]. The formal solution of Eq.(8) can be written as follows \cite{FED,SSH}
\begin{eqnarray}\label{11}
\nonumber
\hat{d}(t)=-(i/\hbar)\lim_{\alpha\rightarrow 0}\int_{0}^{t}dt^{'} \hat{X}^{+}(t^{'})[\gamma_{L} \hat{\Psi}_{-}(0;t^{'})+\nonumber\\
+\gamma_{R} \hat{\Psi}_{-}^{+}(0;t^{'})]e^{-i(\tilde{\Delta}-i\alpha)(t-t^{'})/\hbar}
\end{eqnarray}
(with $\alpha$ being positive infinitesimal). Now, substituting Eq.(11) into Eq.(10), we are getting the following basic equation for the averages
\begin{eqnarray}\label{12}
 \nonumber
  \hbar\lbrace\langle\hat{\Psi}_{-}^{+}(0^{-};t)\hat{\Psi}_{-}(0^{+};t)\rangle-\langle\hat{\Psi}_{-}^{+}(0^{-};t)\hat{\Psi}_{-}(0^{-};t)\rangle\rbrace=
  \nonumber \\
- \lim_{\alpha \rightarrow 0}\int_{0}^{t}dt^{'}\{[\gamma_{L}^{2}\langle\hat{\Psi}_{-}^{+}(0^{-};t)\hat{X}^{+}(t^{'})\hat{X}(t)\hat{\Psi}_{-}(0;t^{'})\rangle
\nonumber \\
+\gamma_{L}\gamma_{R}\langle\hat{\Psi}_{-}^{+}(0^{-};t)\hat{X}^{+}(t^{'})\hat{X}(t)\hat{\Psi}_{-}^{+}(0;t^{'})\rangle]
e^{-i(\tilde{\Delta}-i\alpha)(t-t^{'})/\hbar}\nonumber\\
+[\gamma_{R}^{2}\langle\hat{\Psi}_{-}^{+}(0^{-};t)\hat{X}^{+}(t)\hat{X}(t^{'})\hat{\Psi}_{-}(0;t^{'})
\rangle \nonumber\\
+\gamma_{L}\gamma_{R}\langle\hat{\Psi}_{-}^{+}(0^{-};t)\hat{X}^{+}(t)\hat{X}(t^{'})\hat{\Psi}_{-}^{+}(0;t^{'})
\rangle] e^{i(\tilde{\Delta}+i\alpha)(t-t^{'})/\hbar}\}\nonumber
\end{eqnarray}
\begin{equation}
\label{12}
\end{equation}
here symbol $\langle..\rangle$ stands for the averaging with the total transformed Hamiltonian. 

Now, the central Eq.(12) should be complemented by the corresponding equation for bosonic operator $\hat{p}$. The Heisenberg equation for vibronic subsystem reads
\begin{eqnarray}
\nonumber
[\partial_{t}^{2}+\omega_{0}^{2}]\hat{p}=i\frac{\lambda\omega_{0}}{2\hbar}[\hat{X}^{+}\hat{d}^{+}(\gamma_{L}\hat{\Psi}_{-}(0)+
\gamma_{R}\hat{\Psi}_{-}^{+}(0))\nonumber\\
-\hat{X}\hat{d}(\gamma_{L}\hat{\Psi}_{-}^{+}(0)+
\gamma_{R}\hat{\Psi}_{-}(0))]\nonumber
\end{eqnarray}
\begin{equation}
\label{13}
\end{equation}
and, obviously, it could be rewritten in the form
\begin{equation} \label{14}
[\partial_{t}^{2}+\omega_{0}^{2}]\hat{p}=\frac{\lambda\omega_{0}}{2}\partial_{t}(\hat{d}^{+}\hat{d}).
\end{equation}
To proceed further we need to use certain approximations. The most evident simplification is to put the right-hand side of Eq.(14) to be equal to zero. This corresponds to the case, where bosonic subsystem is effectively unaffected by fermionic one and the Eq.(14) has a free solution
\begin{equation} \label{15}
\hat{p}_{0}(t)=\frac{i}{\sqrt{2}}(\hat{b}_{0}^{+}e^{i\omega_{0}t}-\hat{b}_{0}e^{-i\omega_{0}t}).
\end{equation}
Here the operators $\hat{b}_{0}^{+}$($\hat{b}_{0}$) describe the creation (annihilation) of a free vibron and fulfill standard bosonic commutation relation $[\hat{b}_{0},\hat{b}_{0}^{+}]=1$. One can see from Eqs.(13,14), that the approximation (15) is always valid in the perturbation theory on $\Gamma_{0}=\gamma_{L}^{2}+\gamma_{R}^{2}$ ($\Gamma_{0}\ll \hbar \omega_{0},T,eV$), where the lowest energy scale is $\Gamma_{0}$ -the "bare" width of the fermionic level of QD in the "Wide-Band Approximation" (WBA-limit). Another approach where one can use the Hamiltonian of free vibrons is the so-called "polaron tunneling approximation" \cite{OUR1,NEW}. It is valid, when the characteristic lifetime of electron on QD ($\sim \hbar/\Gamma_{0}$) is much greater than the time of polaron formation ($\sim 1/\lambda^{2}\omega_{0}$). In this case only the polaronic states "live" on the dot and one can use fermion-boson factorization when calculating the Green function of polaron. This approach allows one to consider resonant tunneling in the system with strong electron-vibron interaction ($\lambda \gtrsim 1$). Thus, when evaluating the averages in Eq.(12) we will assume that
\begin{eqnarray}\label{16}
\nonumber
  \langle \hat{\Psi}_{-}^{+}(0^{-};t)\hat{X}^{+}(t^{'})\hat{X}(t)\hat{\Psi}_{-}(0;t^{'}) \rangle \simeq\\
\langle \hat{\Psi}_{-}^{+}(0^{-};t)\hat{\Psi}_{-}(0;t^{'}) \rangle_{\hat{\tilde{H}}_{l}}
\langle \hat{X}^{+}(t^{'})\hat{X}(t)) \rangle_{\hat{\tilde{H}}_{d}}.
\end{eqnarray}
Here the symbols $\langle..\rangle_{\hat{\tilde{H}}_{d}}$ and $\langle..\rangle_{\hat{\tilde{H}}_{l}}$ stand for the averaging with quadratic Hamiltonian of the QD ($\hat{\tilde{H}}_{d}$) and of the leads ($\hat{\tilde{H}}_{l}$). 
In the polaronic approach of Eq.(16) it is natural to regard vibronic subsystem as thermally equilibrated at the temperature $T$. 
\begin{equation} \label{17}
\langle \hat{b}_{0}^{+}\hat{b}_{0}\rangle_{\hat{\tilde{H}}_{d}}=n_{b}(\beta)=(\exp(\beta)-1)^{-1}
\end{equation}
with $\beta=\hbar \omega_{0}/T$.
Obviously, our approximations (15-17) allow one to generalize the scattering approach elaborated in Ref.[\onlinecite{KG1}] for resonant tunneling of interacting electrons to the case of resonant tunneling of polarons. Under the accepted approximations the solution of basic Eq.(12) is formulated in terms of noninteracting $\hat{\Psi}_{-}$ - fermions backscattered from the point $x=0$ or transmitted through it from negative ($x<0$) to positive ($x>0$) half-axis. Following Ref.[\onlinecite{KG1}] we use in Eq.(12) the standard momentum decomposition for fermionic field operator $\hat{\Psi}_{-}(x,t)$
\begin{equation} \label{18}
\hat{\Psi}_{-}(x;t)=\int\frac{dk}{2\pi}e^{ik(t-x)}\left\{
                                                    \begin{array}{ll}
                                                      \hat{a}_{k}, & x<0  \\
                                                      \hat{b}_{k}, & x>0
                                                    \end{array}
                                                  \right.
\end{equation}
(we set here $\hbar=1$ and $v_{g}=v_{F}/g=1$) In Eq.(18) $\hat{a}_{k}^{+}$ ($\hat{a}_{k}$) are the standard fermionic creation (annihilation) operators. It is evident, that the average occupation number $\langle\hat{a}_{k}^{+}\hat{a}_{k}\rangle$ for a new fermionic state at the energy $k$ will be as follows $\langle\hat{a}_{k}^{+}\hat{a}_{k^{'}}\rangle_{\hat{\tilde{H}}_{l}}=[n_{F}(k-\mu_{L})-n_{F}(k-\mu_{R})]\delta(k-k^{'})$,
where $n_{F}(k-\mu_{L(R)})=(\exp(\beta(k-\mu_{L(R)}))+1)^{-1}$ are the Fermi distribution functions of electron in the 
$L(R)$ reservoirs. Analogously to Ref.[\onlinecite{KG1}], we represent operator $\hat{b}_{k}$ in Eq.(18) as $\hat{b}_{k}=t(k)\hat{a}_{k}$, where $t(k)$ is the transmission amplitude. The reflection amplitude $r(k)$ is defined by the equation $\hat{a}_{-k}^{+}=r(k)\hat{a}_{k}$ and describes the process of the Andreev-like reflection of the incoming quasiparticle. We put also $\hat{b}_{-k}^{+}=0$, since there are no transmitted "holes" in the system. The considerations given above allow us to rewrite a basic formula (7) for the average current through our system. It takes the Landauer-type form \cite{KG1,KG2}
\begin{equation}\label{19}
\bar{I}(V)=\langle\hat{I}\rangle=\frac{e}{h}\int d\varepsilon R(\varepsilon)[n_{F}(\varepsilon-eV)-n_{F}(\varepsilon)],
\end{equation}
where $R(\varepsilon)=1-|t(\varepsilon)|^{2}$ is the energy-dependent reflection coefficient for $\hat{\Psi}_{-}$- fermions, which determines the effective transmission coefficient for physical electrons transferred through the QD.

\section{Results and discussion}

One can see that the problem of electron transport in the considered model is reduced now to the evaluation of the effective transmission coefficient $R(\varepsilon)$. Thus, regarding the basic Eq.(12) with its complex conjugated equation, after the averaging under the assumption about a fermion-boson factorization (15)-(18), one can solve the resulting integral equation with respect to the complex transmission amplitude $t(\varepsilon)$ of $\hat{\Psi}_{-}$- fermions. Particularly, after the integrating over $dt^{'}$, taking the limits $t\rightarrow \infty$ and $\alpha\rightarrow 0$, it is possible to derive the following basic formula for the effective transmission coefficient $R(\varepsilon)$ from Eq.(19)
\begin{equation}\nonumber 
 R(\varepsilon)=\frac{4\tilde{\Gamma}^{2}}{[(1+\tilde{B}_{+}^{2})(1+\tilde{B}_{-}^{2})+
 2\tilde{\Gamma}^{2}(1+\tilde{B}_{+}\tilde{B}_{-})+\tilde{\Gamma}^{4}]}.
\end{equation}
\begin{equation}\label{20}
\end{equation}
Here
\begin{equation}\nonumber
\tilde{B}_{\pm}=\tilde{B}_{\pm}(\varepsilon,\beta)=\sum_{l=-\infty}^{+\infty}
\frac{\Gamma_{0} F_{l}(\beta)\gamma_{l\pm}(\varepsilon)}{\Delta_{l}^{2}-\varepsilon^{2}},
\end{equation}
\begin{equation}\label{21}
\tilde{\Gamma}=\tilde{\Gamma}(\varepsilon,\beta)=\sum_{l=-\infty}^{+\infty}
\frac{\Gamma_{0} F_{l}(\beta)\gamma(\varepsilon)}{\Delta_{l}^{2}-\varepsilon^{2}}
\end{equation}
with
\begin{equation}\nonumber
\gamma_{l\pm}(\varepsilon)=\frac{1}{2}\left[\left( \frac{\gamma_{L}^{2}-\gamma_{R}^{2}}{\gamma_{L}^{2}+\gamma_{R}^{2}}\right)\Delta_{l}\pm \varepsilon \right],
\end{equation}
\begin{equation}\label{22}
\gamma(\varepsilon)=\left(\frac{\gamma_{L}\gamma_{R}}{\gamma_{L}^{2}+\gamma_{R}^{2}}\right)\varepsilon,
\end{equation}
and
\begin{equation}
F_{l}(\beta)=e^{-\lambda^{2}(1+2n_{b})}I_{l}(\lambda^{2}\sqrt{n_{b}(1+n_{b})})e^{-\beta l/2}.
\label{23}
\end{equation}
In Eqs.(21,23) $I_{l}(z)$ is the Bessel function of $l$-th order of the imaginary argument. In Eqs.(22) $\Delta_{l}=\tilde{\Delta}-\hbar\omega_{0}l$ is the energy of the $l$-th resonance. 

Formulas (20)-(23) generalize the results obtained by Komnik and Gogolin in Ref.[\onlinecite{KG1}] on the case of strong electron-vibron interaction in the QD and represent the basic result of this paper. Indeed, in the case when $\lambda=0$ (i.e. in the absence of electron-vibron interaction) our general formula (20) is reduced to the basic result (Eq.(10)) of Ref.[\onlinecite{KG1}] for the general case of asymmetric tunnel barriers ($\gamma_{L}\neq\gamma_{R}$). 
One can see that the influence of the asymmetry of tunnel barriers on the transport properties of the system is concerned mostly in the renormalization $\sim \eta R_{S}(\varepsilon)$ of the effective transmission coefficient $R_{S}(\varepsilon)$ for the symmetric junction ($\gamma_{L}=\gamma_{R}$). Here $\eta\leq1$ is the asymmetry parameter $\eta=(\gamma_{L}/\gamma_{R})^{2}$ if $\gamma_{L}<\gamma_{R}$, and $\eta=(\gamma_{R}/\gamma_{L})^{2}$ if $\gamma_{R}<\gamma_{L}$. 
\begin{figure}
\includegraphics[height=10 cm,width=8.6 cm]{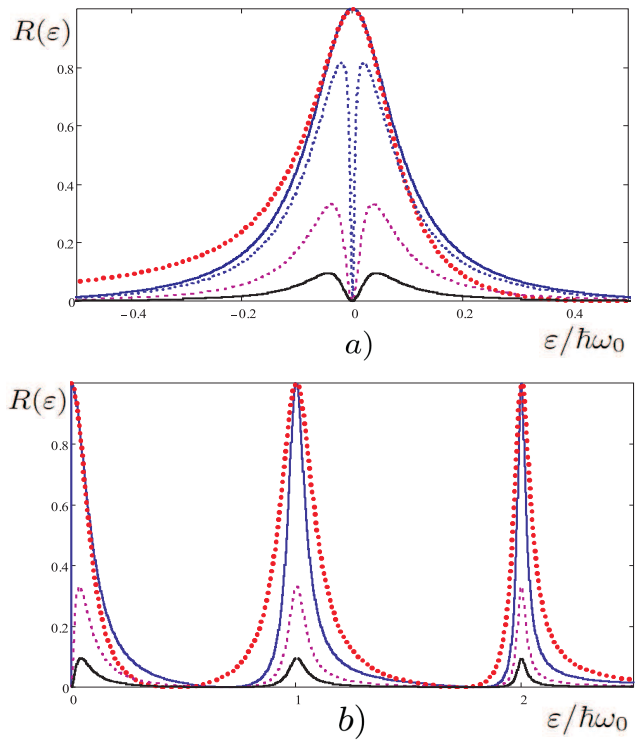}
\caption{The effective transmission coefficient $R(\varepsilon)$ of Eq.(20) as the function of energy (in the units of $\hbar \omega_{0}$), for different values of the asymmetry parameter $\eta=\Gamma_{L(R)}/\Gamma_{R(L)}$, in comparison with effective transmission coefficient (red dotted line) obtained in Ref.[\onlinecite{NEW}] (Eqs.(22,23)) for the $g=1$- model with symmetric tunnel barriers ($\eta=1$). On Fig.1a zero-bias peak is depicted for: $\eta=1$ (symmetric case) - blue solid line; $\eta=0.4$ - blue dotted line;  $\eta=0.1$ - magenta dotted line; and  $\eta=0.025$ (strongly asymmetric case) - black solid line. On Fig.1b, in addition to the zero-bias peak, two "polaron-assisted" peaks are shown. These peaks describe resonant Andreev-like polaron-assisted tunneling. Here $\eta=1$ (symmetric case) - blue solid line; $\eta=0.1$ - magenta dotted line; and $\eta=0.025$ (strongly asymmetric case) - black solid line. On Figs.1a,b we also put $\lambda^{2}=1$,  $\Gamma_{0}/\hbar\omega_{0}=0.25$.}
\end{figure}

In the asymmetric case, as one can see from Fig.1a, the "zero-bias" (with $l=0$) resonance has a dip at $\varepsilon\rightarrow0$, if $\tilde{\Delta}=0$. This dip shrinks to zero, when $\varepsilon=0$ at the arbitrary value of asymmetry parameter $\eta<1$. The presence of such dip at $\varepsilon\rightarrow 0$ in the case of even small asymmetry of tunnel barriers is the manifestation of the Kane-Fisher effect \cite{KF,FUR} at $\varepsilon\rightarrow \varepsilon_{F}=0$. This is because at $\eta\neq1$ the specific particle-hole symmetry in our model breaks down, and more common Luttinger liquid physics is revealed. 

Thus, all distinguishing features of the model can be obtained by considering a more simple case of symmetric tunnel barriers $\gamma_{L}=\gamma_{R}$. In this case our central expression (20) for the effective transmission coefficient $R_{S}(\varepsilon)$ takes a very transparent form
\begin{equation} \label{24}
 R_{S}(\varepsilon)=\frac{1}{1+(S(\varepsilon))^{-2}}
\end{equation}
with
\begin{equation}\label{25}
S(\varepsilon)=\sum_{l=-\infty}^{+\infty}\frac{\varepsilon \Gamma_{0}F_{l}(\beta)}{\Delta_{l}^{2}-\varepsilon^{2}}.
\end{equation}

Evidently, in the absence of electron-vibron interaction ($\lambda=0$), formulae (24,25) reproduce the result for resonant tunneling between Luttinger liquid leads with $g=1/2$ in the symmetric case \cite{NG,KG1,KG2}. Particularly, at $\tilde{\Delta}=0$ and $\lambda=0$, Eq.(24) turns into the usual Breit-Wigner expression for transmission coefficient of noninteracting electrons. This is because, in that case the mapping on the model with Fermi leads ($g=1$) becomes valid \cite{KG1}. 

Now let us analyze formulae (24),(25) in a more general situation, where $\lambda\neq0$. It is worth to point out, that by means of the physically transparent method formulated above, one could also solve the model with Fermi- leads ($g=1$- case) for strong electron-vibron interaction $\lambda\geq 1$ (see Ref.[\onlinecite{NEW}]). In particular, for the symmetric junction ($\gamma_{L}=\gamma_{R}=\gamma_{0}$), eliminating the exotic coupling terms $\hat{d}^{+}\hat{X}^{+}\gamma_{0} \hat{\Psi}_{-}^{+}(0)$ and $\gamma_{0} \hat{\Psi}_{-}(0)\hat{X}\hat{d}$ from the tunnel Hamiltonian (6), and applying the entire scheme described above, one could reproduce the central results (Eqs.(22),(23)) of Ref.[\onlinecite{NEW}], obtained there by means of more rigorous mathematical methods (Full Counting Statistics (FCS) and Keldysh technique). This is because, the central approach of Ref.[\onlinecite{NEW}] (the linear diagram resummation scheme) is included into our polaronic approximation of Eqs.(15-18). 

One can see, that in general case (if $\lambda\neq0$ and $\Delta_{l}\neq0$), the expressions (24-25) differ from corresponding formulae (Eqs.(22-23)) of Ref.[\onlinecite{NEW}] for the case of Fermi leads (with $g=1$). Indeed, in our model a special symmetry: $\hat{\rho}_{L(R)}(x)\leftrightarrow -\hat{\rho}_{R(L)}(x)$ of charge density excitations on different physical leads takes place. Such symmetry makes electrons from different leads be strongly correlated in the vicinity of QD and implies a special (Andreev-like) type of resonant tunneling in the system at $\Delta_{l}\neq 0$ and $\gamma_{L}=\gamma_{R}$. In the latter case, a perfect transmission of the electron through QD, along the infinite one-dimensional system (when in Eq.(19) one has $R_{S}(\varepsilon)=1$), can be treated as the total Andreev reflection of spatially nonlocal $\hat{\Psi}_{-}$- fermion from the boundary at the point $x=0$ back to the half-infinite 1D system with $x<0$. During such a reflection the incoming particle-type excitations are transformed into the opposite-moving excitations of the hole-type at the same energies. 

The energy-dependent effective transmission coefficient of Eq.(20) is plotted on Fig.1 for different values of the asymmetry parameter $\eta=\Gamma_{L(R)}/\Gamma_{R(L)}$, in comparison with one, calculated in Ref.[\onlinecite{NEW}] for the case of  Fermi leads ($g=1$) and symmetric tunnel barriers. It is essential, that in the resonant tunneling limit at $T=0$ all obtained from Eqs.(20,24) resonant peaks describe elastic processes of a perfect transmission of the polaron at resonant energies $\varepsilon_{l}=\hbar \omega_{0}l$, $l=0,\pm1,2,..$. In such processes, one or more virtual vibrons are emitted and, then absorbed, leaving the fermionic subsystem of the QD in the same quantum state as before the polaron transmission \cite{NEW}. On the other hand, in the opposite case of sequential tunneling ($\Gamma_{L,R}\ll T \ll \hbar\omega_{0}$) only the inelastic tunneling, accompanied by the emission of real vibrons, produce small (satellite) resonances, while all virtual processes result only in the "polaronic narrowing" of all resonances \cite{MCK,FL}. 

Regarding the off-resonant energies: $\hbar \omega_{0}\gg \vert\varepsilon_{l}-\Delta_{l}\vert\gg \Gamma_{0}$, one may conclude from Fig.1b, that the transmission coefficient (20) shrinks to zero at these energies if $\lambda^{2}\Gamma_{0}\gg T\rightarrow0 $. This is the consequence of destructive interference of different virtual polaronic states on the QD \cite{NEW}. As one can see from Fig.1, although a perfect transmission at zero temperature in both cases (for $g=1/2$- and $g=1$- models) takes place at the same energies $\varepsilon_{l}=\hbar \omega_{0}l$, $l=0,\pm1,2,..$, the electron-electron interaction in the $g=1/2$- case sufficiently narrows all "polaron-assisted" (with $l\geq1$) resonances, as compared to the case of Fermi leads ($g=1$). This fact clearly shows, that although a resonant polaron-assisted tunneling is possible for both systems with $g=1$- and $g=1/2$- leads, the details of such processes are different in these two cases. Indeed, the resonant tunneling in $g=1/2$- case requires from the electrons placed on different leads to be in definite strongly correlated quantum states during the process of resonant tunneling. These strong correlations entangle physical electrons from different leads in the vicinity of QD. Obviously, the probability of resonant state, which involves more than one physical electron (when $g=1/2$) at resonant energy far from $\varepsilon_{F}=0$ (i.e. in the case $l\geq1$) is expected to be much smaller than the probability of the resonant state, which involves only a single electron of the same energy (when $g=1$). To estimate this effect quantitatively for the case $\Gamma_{0}\ll\hbar\omega_{0}\ll eV$, one could perform the effective transmission coefficient of Eq.(24) at resonant energy $\varepsilon_{l}$ near the $l$-th polaron-assisted resonance ($l\geq1$) as follows
\begin{equation} \label{26}
 R_{S0}(\varepsilon_{l})\simeq \frac{\Gamma_{eff(l)}^{2}}{\Gamma_{eff(l)}^{2}+(\Delta_{l}-\varepsilon_{l})^{2}}
\end{equation}
with the following approximate expression for the effective width $\Gamma_{eff(l)}$ of each $l$-th ($l=1,2,..,l_{m}$) resonance
\begin{equation} \label{27}
\Gamma_{eff(l)}\approx\frac{\Gamma_{l}/2}{1+\Gamma_{l}/(2\Delta_{l})}.
\end{equation}
Here, 
\begin{equation} \label{28}
\Gamma_{l}=\Gamma_{0}e^{-\lambda^{2}}(\lambda^{2})^{l}/l!
\end{equation}
-is a low-temperature asymptotic for the effective width of the $l$-th resonance in the case of Fermi leads \cite{NEW} and $\Delta_{l}=\hbar\omega_{0}l$, ($\tilde{\Delta}=0$), $0\leq l \leq l_{m}$, with the maximal number of vibrons emitted $l_{m}\simeq [eV/\hbar\omega_{0}]$. This is because, at $T,\Gamma_{0}\ll \hbar\omega_{0}$ there are no processes of absorption of any "external" vibrons and since the energies of incoming quasiparticles are limited by $eV$ at $T \ll eV$. 

The important consequence of Eqs.(27,28) is that, the obtained relative additional narrowing $\Gamma_{eff(l)}/\Gamma_{l}$ due to electron-electron interaction is found to be strong ($\sim1/2$) and approximately the same for all polaron-assisted resonances. This feature seems to be specific only for the considered model and might help us in the distinguishing of this important $g=1/2$- case among other experimental realizations of 1D molecular transistors. This is clearly illustrated by Fig.2, where the effective widths of polaron-assisted resonances are plotted (in the units of $\Gamma_{0}$), as functions of electron-vibron interaction $\lambda^{2}$ for different values of $l\geq1$ in both $g=1/2$- and $g=1$- cases. 
\begin{figure}
\includegraphics[height=6 cm,width=8.6 cm]{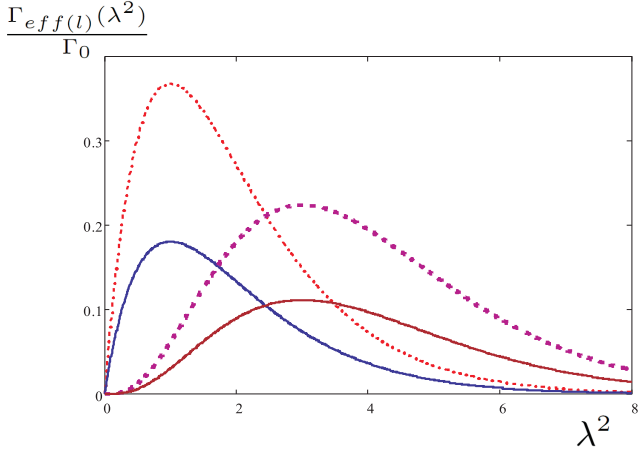}
\caption{The effective widths of different polaron-assisted resonances from Eq.(27)(in the units of $\Gamma_{0}$) for symmetric junction ($\eta=1$), as the functions of electron-vibron interaction $\lambda^{2}$. Blue solid line corresponds to effective width of polaron-assisted resonance for $g=1/2$-model with $l=1$, while the brown solid line represents the resonance with $l=3$ for the same model. Red dotted line and magenta dotted line correspond to widths of polaron-assisted resonances with $l=1$ and $l=3$, obtained for the $g=1$-model \cite{NEW}. (Here we put $\Gamma_{0}/\hbar \omega_{0}=0.1$.)}
\end{figure}

From Fig.2 one can easily observe two main effects. First of all, one can see the described above effect of the additional narrowing of all polaron-assisted resonances, due to specific electron-electron interaction in the system with $g=1/2$- LL leads. As it could be seen from Eq.(27), this effect is strong enough at $l\geq1$. Besides that, for strong electron-vibron interaction (if $\lambda\geq 1$) in both $g=1/2$- and $g=1$- cases a strongly non-monotonic behaviour of the effective resonance widths (with the maximum at $\lambda^{2}\simeq l$) as functions of $\lambda^{2}$ takes place. The latter phenomenon is a well-known consequence of the fermion-boson factorization procedure and is caused by the interplay of the effects of vibron-assisted tunneling and polaronic blockade \cite{OUR,NEW}. Although our general formulae are valid in the limiting case $\lambda=0$ for both resonant and off-resonant tunneling, for $\lambda\rightarrow0$ ($\lambda\neq 0$), the predictions of our method coincide with the results of perturbation theory calculations in small $\lambda^{2}$ for each $l\neq 0$ only while $T/\Gamma_{0}\gg(\lambda^{2}/l)^{l}$. Thus, it is reasonable to conclude, that if $\lambda\ll1$ (but $\lambda\neq 0$) the case $T/\Gamma_{0}\ll(\lambda^{2}/l)^{l}\ll 1$ ($\lambda\rightarrow0$, $T=0$) is not described by means of our method \cite{K}.

Finally, let us consider the behaviour of the differential conductance $G_{S0}(V)=d\bar{I}(V)/dV$ for our model in the simplest symmetric case ($\gamma_{L}=\gamma_{R}$). At high temperatures $\lambda^{2}\hbar\omega_{0}\ll T$, when all thermally activated (vibronic) channels contribute to electron tunneling, the polaronic blockade in the zero-bias peak is totally lifted \cite{KN,UM} and we get a standard high-temperature asymptotic $\sim G_{0}\Gamma_{0}/T$ for the conductance of the $g=1$-model with noninteracting electrons (here $G_{0}=e^{2}/h$ is the conductance quantum). This is because, at high temperatures all specific quantum features of both electron-electron and electron-vibron interactions are "smeared out" by thermal fluctuations. Much more interesting is the case of low temperatures, where $T\ll\hbar\omega_{0}$. In this case, a main contribution to effective transmission coefficient (24) goes from the resonant terms of the form (26). Therefore, one could obtain a following low-temperature asymptotic for differential conductance $G_{S0}(V)$
\begin{eqnarray}\label{29}
\nonumber
G_{S0}(V)=G_{0}\sum_{l=0}^{l_{m}}\frac{1}{2}\lbrace\tanh\left(\frac{\Delta_{l}-eV+\Gamma_{eff(l)}/2}{2T}\right)-\nonumber\\
-\tanh\left(\frac{\Delta_{l}-eV-\Gamma_{eff(l)}/2}{2T}\right)\rbrace\nonumber
\end{eqnarray}
\begin{equation}\label{29}
\end{equation}
In Eq.(29) $\Delta_{l}=\hbar\omega_{0}l$ is the energy of the $l$-th resonant level ($l=0,1,2,..,l_{m}$) (-here the "detuning" term $\tilde{\Delta}$ is insufficient, due to the existence of the gate voltage, thus we put it to be equal to zero). The effective widths $\Gamma_{eff(l)}$ are from Eq.(27), and the maximal number $l_{m}$ of vibrons emitted is of the order of the integer part of quantity $[eV/\hbar\omega_{0}]$. The resulting differential conductance (29), as the function of bias $V$, exhibits a sequence of sharp resonances at $eV=eV_{l_{m}}=\Delta_{l_{m}}$ ($l_{m}=0,1,2,..$). The main contribution to the sum in Eq.(29) goes from the resonant term with $l_{m}\simeq [eV/\hbar\omega_{0}]$ if 
$T\ll \hbar\omega_{0}\ll eV$, and from the term with $l_{m}=0$ if $T \ll eV \ll \hbar\omega_{0}$, correspondingly. Thus, differential conductance (29), as the function of resonant values of bias $eV_{lm}\simeq \Delta_{l_{m}}$ can be estimated for these two cases as follows
\begin{equation}\label{30} \nonumber
G_{S0}(V_{lm})\approx G_{0}\left\{
\begin{array}{ll}
\tanh\left[\frac{\Gamma_{0}}{4T}e^{|\frac{eV_{lm}}{\hbar\omega_{0}}|\ln\left|
\frac{\lambda^{2}\hbar\omega_{0}}{eV_{lm}}\right|-\lambda^{2}}\right], \\ &
\\
(T\ll \hbar\omega_{0}\ll eV_{lm}\simeq\hbar\omega_{0}l_{m});  \\
\\
\tanh\left[\frac{\Gamma_{0}e^{-\lambda^{2}}}{4T}\right]_{l_{m}=0}, \\ &
\\ (T\ll eV_{lm}\ll \hbar\omega_{0}); \\
\end{array}
\right.
\end{equation}
\begin{equation}
\label{30}
\end{equation}

Evidently, the Eq.(30) describes both resonant (where $T\ll\Gamma_{eff(l_{m})}$), and sequential (where $T\gg\Gamma_{eff(l_{m})}$) regimes of polaron tunneling as well. Indeed, if $eV_{lm}=eV_{0}\ll \hbar\omega_{0}$ ($l_{m}=0$) then $G_{S0}(V_{0})\sim G_{0}\Gamma_{0}e^{-\lambda^{2}}/T$. This formula reproduces a well-known \cite{MCK,FL,KO,OUR} $\Gamma_{\lambda}/T$-scaling for the low-temperature conductance of the $g=1$-model, with the renormalized bare level width $\Gamma_{\lambda}=\Gamma_{0}e^{-\lambda^{2}}$, as the result of polaronic (Frank-Condon) blockade in the elastic channel of tunneling. For differential conductance ($eV_{lm}\gg\hbar\omega_{0}$) in the sequential tunneling limit 
$T\gg(1/2)(\lambda^{2}/l_{m})^{l_{m}}\Gamma_{0}e^{-\lambda^{2}}$, ($l_{m}\simeq[eV_{lm}/\hbar\omega_{0}]\gg 1$ and 
$T\ll \hbar\omega_{0}\ll eV_{lm}$) the situation is quite similar. The expected $1/T$-scaling of differential conductance (30) in this case will be the following
\begin{equation} \nonumber
G_{S0}(V_{lm})\approx
G_{0}\left(\frac{1}{2}\right)\frac{\Gamma_{0}e^{-\lambda^{2}}}{T}
\left(\frac{\lambda^{2}\hbar\omega_{0}}{eV_{lm}}\right)^{\frac{eV_{lm}}{\hbar\omega_{0}}}_{eV_{lm}\simeq\hbar\omega_{0}l_{m}}.
\end{equation}
\begin{equation}\label{31}
\end{equation}
Formula (31) gives us the "hights" of low-temperature differential conductance peaks in the sequential tunneling limit, at resonant values of bias: $eV_{lm}\simeq \hbar\omega_{0}l_{m}$, $l_{m}=1,2,..$ in the case of symmetric junction. The only difference between expression (31) and corresponding limiting case for the $g=1$- model is a pre-factor $\sim 1/2$ in Eq.(31). It results in a sufficient suppression of all satellite (with $l_{m}\geq 1$) peaks, as compared to the $g=1$- case. Naturally, the above expansion of Eq.(30) for the sequential tunneling limit coincides (up to a pre-factor $\sim 1/2$) with the result of perturbation theory for the $g=1$- model in small $\Gamma_{0}$ ($\Gamma_{0}\ll T$), or in small $\lambda^{2}$ at non-zero temperatures (if $\lambda^{2}\ll T/\Gamma_{0}$) \cite{MCK,FL}. Naturally, this fact is because of the mapping on the $g=1$- model. A novel feature here is that, such mapping is not complete any more for energies 
$\varepsilon \simeq eV \geq \hbar\omega_{0}$, even in the sequential tunneling limit. Only for $eV\ll \hbar\omega_{0}$ the correspondence between the $g=1/2$- and $g=1$- models with strong electron-vibron interaction is complete. Of course, the standard perturbation theory in small $\Gamma_{0}$ for the SET weakly coupled to the LL-leads with arbitrary $g$  \cite{FUR,OUR,CHINA} ($0<g\leq 1$) is unable to give a correct description for our case, since it does not take into account strong correlations between tunneling events, which are specific for $g=1/2$- system. 
\begin{figure}
\includegraphics[height=6 cm,width=8.6 cm]{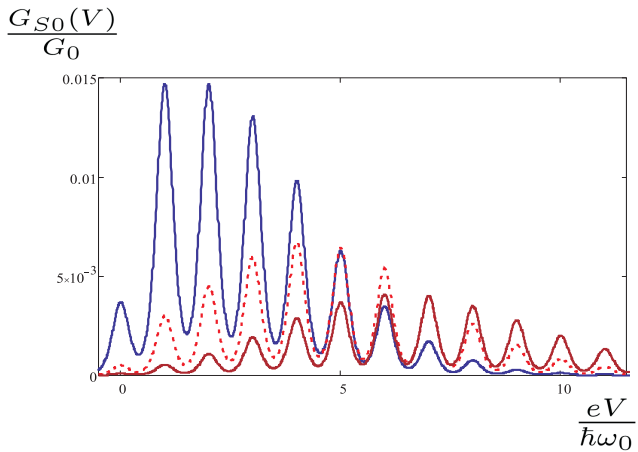}
\caption{The low-temperature differential conductance $G_{S0}(V)$ (in the units of conductance quantum $G_{0}$) from Eq.(29), as the function of bias $V$ (in the units of $\hbar \omega_{0}/e$), for different values of electron-vibron interaction $\lambda^{2}$. Here the blue solid line represents the case $\lambda^{2}=4$, while the red dashed line corresponds to $\lambda^{2}=6$, and the brown solid line describes the case $\lambda^{2}=8$ (-strong electron-vibron interaction). Also we put here $\hbar \omega_{0}/T=8$ and $\Gamma_{0}/\hbar \omega_{0}=0.1$.} 
\end{figure}

As regards to the resonant tunneling limit $T\ll\Gamma_{eff(l_{m})}$ in Eqs.(29-30), the interplay of the effects of electron-electron correlations and strong electron-vibron interaction becomes nonperturbative in this case. In particular, in this limit there is no significant difference between the peaks of differential conductance, corresponded to polaron-assisted (with $l_{m}\neq0$) and zero-bias (with $l_{m}=0$) channels of tunneling. Besides that, strong interactions of both types (electron-electron and electron-vibron) lead to a sufficient narrowing of all polaron-assisted (with $l\neq0$) resonances at: $\lambda^{2}\geq 1$, whereas in the zero-bias channel (with $l=0$) only a polaronic narrowing takes place at: $\lambda\geq1$. In the zero-temperature limit $T\rightarrow 0$, one will have a perfect resonant tunneling of a polaron with conductance quantum $G_{S0}(V_{lm})\rightarrow G_{0}$ at resonance values of bias $eV_{lm}=\Delta_{l_{m}}$, ($l_{m}=1,2,..$), but only if $0\leq T/\Gamma_{0}\leq (\lambda^{2}/l_{m})^{l_{m}}$, ($T\rightarrow 0$). Thus, evidently, the zero-temperature perturbation theory in small $\lambda^{2}$ (see Ref.[\onlinecite{K}]) does not describe the considered case of resonant polaron-assisted tunneling in the $g=1/2$- model at $T=0$. 

On the other hand, in the case of strong electron-vibron interaction ($\lambda\geq 1$) at low, but nonzero temperatures ($0<T/\Gamma_{0}\ll 1$) as well, as in the discussed above sequential tunneling limit, one can observe in the model the domination of vibron-assisted electron transport (see Eqs.(29-30) and Fig.3). This effect was predicted earlier \cite{OUR,CHINA}, but only in the limits of perturbation theory in $\Gamma_{0}$. On Fig.3 the differential conductance of Eq.(29) is plotted (in the units of $G_{0}$), as the function of bias voltage $V$ for the "intermediate" region of temperatures, between the "resonant" and the "sequential" tunneling regimes (when $T/\Gamma_{0}\gtrsim 1$), in the case of strong electron-vibron interaction ($\lambda\geq1$). One can see from Fig.3, that polaron-assisted (with $l\neq0$) resonant peaks strongly dominate the zero-bias (with $l=0$) peak for all values of $\lambda\geq1$. Therefore, the highest peak on Fig.3 is always polaron-assisted and corresponds to the case where $eV=eV_{lm}\simeq \lambda^{2}\hbar \omega_{0}$, for every given value of $\lambda^{2}$ ($\lambda\geq1$). 

In the zero-temperature limit, the resonant average current through the system will be the following
\begin{equation} \nonumber
\bar{I}_{0}(V_{lm})\approx 
\frac{e}{h}\Gamma_{0}e^{-\lambda^{2}}\left[\sum_{l=0}^{l_{m}(V_{lm})}\frac{(\lambda^{2})^{l}}{l!}\right ]_{l_{m}\simeq[eV_{lm}/\hbar\omega_{0}]},
\end{equation}  
\begin{equation}
\label{32}
\end{equation}
with the bias-dependent effective width of the highest virtual resonant level. It is evident from Eq.(32), that zero-temperature average current as the function of bias reaches its maximal value $\bar{I}_{0max}=(e/h)\Gamma_{0}$ in the limit $eV/\hbar\omega_{0}\rightarrow\infty$, when all possible bias-activated virtual channels of polaron-assisted tunneling are opened and, as a result, the polaronic blockade is totally lifted \cite{KN,UM}. Thus, it is reasonable to conclude that, in the case of symmetric junction, the additional narrowing of Eq.(27), as well as the existence of perfect transmission of a polaron at sequence of resonance energies $\hbar \omega_{0}l$, $l=0,1,2,..$ in the zero-temperature limit ($T\rightarrow0$), may serve as the manifestation of the novel (Andreev-like) type of polaron-assisted resonant tunneling of strongly interacting electrons. We think, that the resonant tunneling of such a type is unique for the considered $g=1/2$-model with strong electron-vibron interaction and represents a consequence of strong correlations between the electrons from different physical leads of the system.

\section{Summary}

In the above, the resonant tunneling of strongly interacting electrons through a single-level vibrating quantum dot (QD) is considered for the case of strong electron-vibron interaction in the QD. Corresponding transport problem is solved in terms of the scattering of noninteracting fermions, which involve the "entangled" electrons from both physical leads of the system in the vicinity of QD. As a result, the general formulae for the effective transmission coefficient and for differential conductance are obtained. It is found, that in the case of symmetric junction, for strong electron-vibron interaction, at sufficiently low and zero temperatures, a novel (Andreev-like) type of resonant polaron-assisted tunneling is realized. It turns out, that some features of this type of tunneling are quite similar to ones for the noninteracting electrons (in the case of Fermi leads, where $g=1$). Especially, such resonant tunneling is characterized at zero temperature by perfect transmission (with conductance quantum) of a polaron at sequence of resonance voltages 
$eV_{lm}\approx\hbar \omega_{0}l_{m}$, $l_{m}=0,1,2,..$. The effective widths of all polaron-assisted (with $l_{m}\neq 0$) resonances depend non-monotonically on electron-vibron interaction constant $\lambda$, with the maximum at $\lambda^{2}\simeq l_{m}$ for each $l_{m}$-th ($l_{m}\neq 0$) resonance. This feature leads to the domination of polaron-assisted electron transport in the case of strong electron-vibron interaction. But, despite partial mapping on the noninteracting ($g=1$) case, the most important difference between resonant tunneling in both $g=1/2$- and $g=1$- models is the additional narrowing of the widths of all polaron-assisted resonances in the $g=1/2$- model, as compared to the $g=1$- case. This relative additional narrowing is found to be strong and roughly the same ($\sim1/2$) for all polaron-assisted resonances with: $l_{m}\geq1$. 

Such novel feature points out on special mechanism of polaron-assisted resonant tunneling, which seems to be unique for the considered $g=1/2$- model. Particularly, in the case of $g=1/2$- LL leads, the resonant quantum state of certain energy represents the "entangled" quantum state of the electrons from both physical leads of the system. As the consequence, physical electrons from different leads of the system become strongly correlated, due to the special type of electron-electron interaction. It is relevant, that the revealed additional narrowing (roughly in $1/2$ times) of all polaron-assisted resonances, as compared to the case of Fermi leads, may serve, as the important distinguishing feature of such special $g=1/2$-type of 1D electron systems. 

In conclusion, the author would like to thank I.V.Krive, and also to S.I.Kulinich, R.I.Shekhter and L.A.Pastur for the valuable discussions.

\end{document}